\documentclass[aps,twocolumn,superscriptaddress]{revtex4}
\usepackage{times}
\usepackage{graphicx,graphics,supertabular,fancyhdr}
\usepackage{color}
\usepackage{siunitx}
\usepackage{amsmath}

\newcommand{\bvec}[1]{\ensuremath{\mathbf{#1}}}

\begin{document}

\title{{Kolmogorovian active turbulence of a sparse assembly of interacting swimmers}}

\author{{Mickael Bourgoin}}
\affiliation{Laboratoire de Physique, Univ Lyon, ENS de Lyon, Univ Lyon 1, CNRS, F-69342 Lyon, France}

\author{Ronan Kervil}
\affiliation{ILM, Univ Lyon, Univ Lyon 1, CNRS, F-69622 Villeurbanne CEDEX, France}

\author{Cecile Cottin-Bizonne}
\affiliation{ILM, Univ Lyon, Univ Lyon 1, CNRS, F-69622 Villeurbanne CEDEX, France}

\author{Florence Raynal}
\affiliation{LMFA, Univ Lyon, Centrale Lyon, INSA Lyon, Univ Lyon 1, CNRS, F-69134 \'Ecully, France}

\author{Romain Volk}
\affiliation{Laboratoire de Physique, Univ Lyon, ENS de Lyon, Univ Lyon 1, CNRS, F-69342 Lyon, France}

\author{{Christophe Ybert}}
\affiliation{ILM, Univ Lyon, Univ Lyon 1, CNRS, F-69622 Villeurbanne CEDEX, France}


%
\begin{abstract}
Active matter, composed of self-propelled entities, forms a wide class of out-of-equi\-librium systems that display striking collective behaviors among which the so-called \emph{active turbulence} where spatially and time disordered flow patterns spontaneously arise in a variety of {active systems}. 
De facto, the \emph{active turbulence} naming suggests a connection with a second seminal class of out-of-equilibrium systems, fluid turbulence, and yet of very different nature with energy injected at global system scale rather than at the elementary scale of single constituents. 
Indeed the existence of a possible strong-tie between active and canonical turbulence remains an open question and a field of profuse research. Using an assembly of  self-propelled interfacial particles, we show experimentally that this active system shares remarkable quantitative similarities with canonical fluid turbulence, as described by the celebrated 1941 phenomenology of Kolmogorov. Making active matter entering into the universality class of fluid turbulence not only benefits to its future development but may also provide new insights for the longstanding description of turbulent flows, arguably one of the biggest remaining mysteries in classical physics.
\end{abstract}

\date{This manuscript was compiled on \today}


\maketitle

\section{Introduction}

Active living organisms, such as bacterial suspensions, birds, fishes, etc., tend to self-organize (in swarms, schools, flocks, etc.) and to develop coherent collective motions, with important consequences in terms for instance of nutrient finding strategies or protection against predators~\cite{bib:dombrowski2004_PRL, bib:wensink2012_PNAS, bib:marchetti2013_RevModPhys, bib:doostmohammadi2017_NatureCom, bib:guillamat2017_NatureCom, bib:wu2017_Science}. Such systems share the feature of being intrinsically out-of-equilibirium, as energy is constantly injected at the level of each individual entity. This makes their statistical modeling 
a conceptual challenge, as usual tools from statistical thermodynamics at equilibrium become caducous. Interestingly, the emergence of large scale collective dynamics while energy sources are at small scale, underlies the existence of multi-scale correlations, driven by inter-entities interactions. 

This scenario naturally resonates with the energy cascade phenomenology, the cornerstone of fluid turbulence description. 
A \emph{direct} turbulent energy cascade describes the process where mechanical energy injected into a flow at some large scale $L$ (e.g. by steering of shearing the fluid) flows down to smaller scales (generating intricate multi-scale motions) until it is dissipated into heat by viscous friction at some small scale $\eta$.  
Such a \emph{direct} cascade drives the multi-scale dynamics of 3D turbulence (2D turbulence exhibits an \emph{inverse} cascade, where energy flows from small to large scales).

%
In 1941 Kolmogorov proposed a self-similar description of this process~\cite{bib:K41}.
{He} predicted that there exists an inertial range {of length-scales in between a dissipative small-scale  $\eta$ and a large forcing scale $L$,} for which the statistical moments of velocity increments depend only on the scale $r$ and energy dissipation per unit mass $\epsilon$. Translated into Fourier space, this gave the first quantitative interpretation of the well-known $k^{-5/3}$ energy spectrum of fluid turbulence, where $k=2\pi/r$ is the wave number \footnote{{In the framework of Kolmogorov's theory, Eulerian velocity structure functions, defined as the statistical moments of the velocity increments $\delta_r\bvec{u} = \bvec{u}\left(\bvec{x}+\bvec{r}\right)-\bvec{u}\left(\bvec{x}\right)$ between points of the flow separated by a distance $r=|\bvec{r}|$, are expected to scale as $S_n(r)=\left<\left|\delta_r \bvec{u}\right|^n\right>\propto \left(\epsilon r\right)^{n/3}$ for $r$ within the inertial range. The $k^{-5/3}$ energy spectrum of fluid turbulence is the Fourier space equivalent to $S_2(r)\propto\left(\epsilon r\right)^{2/3}$, with $k=2\pi/r$.}}. Although Kolmogorov's approach has been rapidly shown to fail predicting high order moments (typically $n>4$) due to intermittency~\cite{bib:frisch}, it has so influenced the field that it is referred to as \emph{K41 phenomenology}.
%

%
%
\begin{figure*}[t]
	\begin{center}
		
		 \includegraphics[height=6cm]{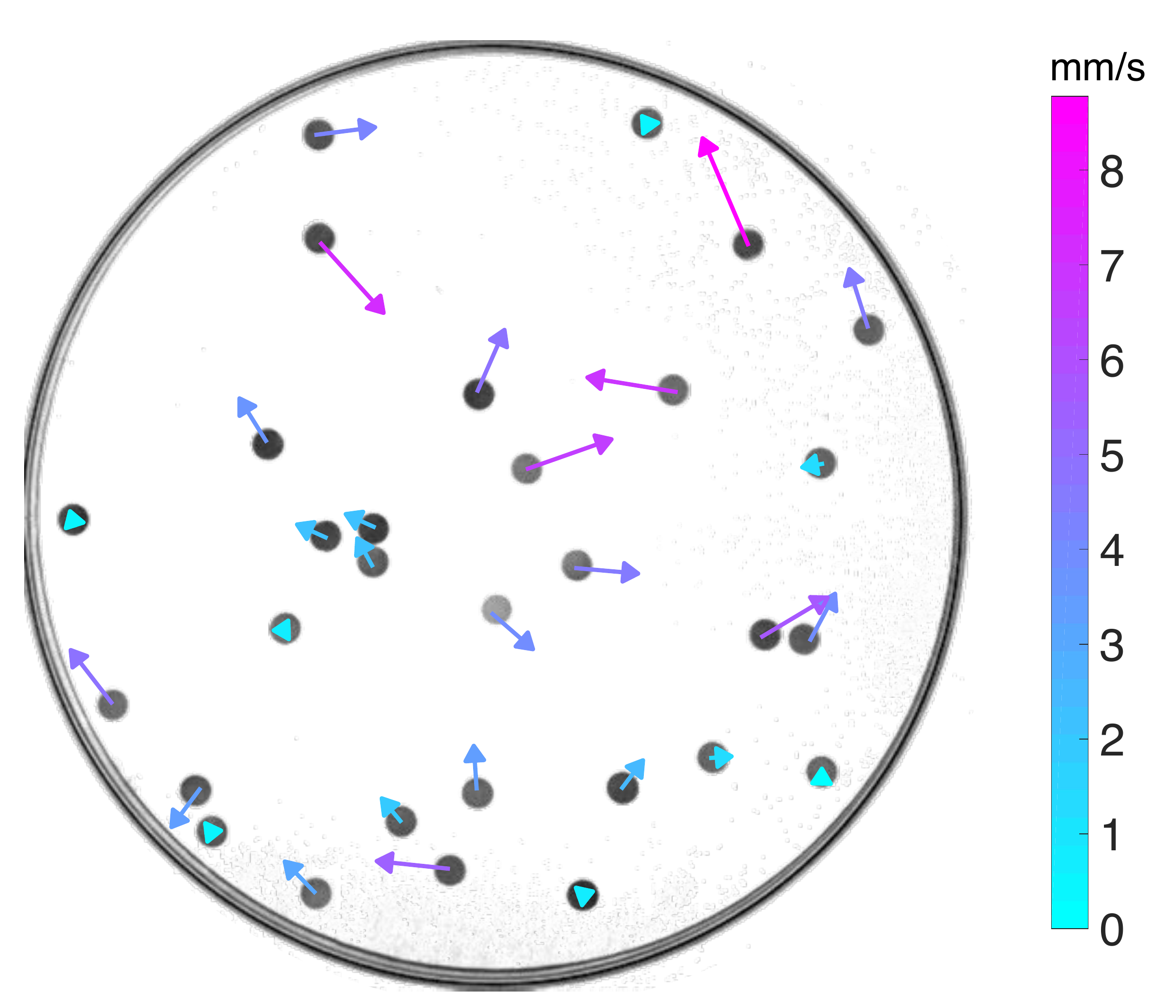}
		 \hspace{1cm}
		\includegraphics[height=6cm]{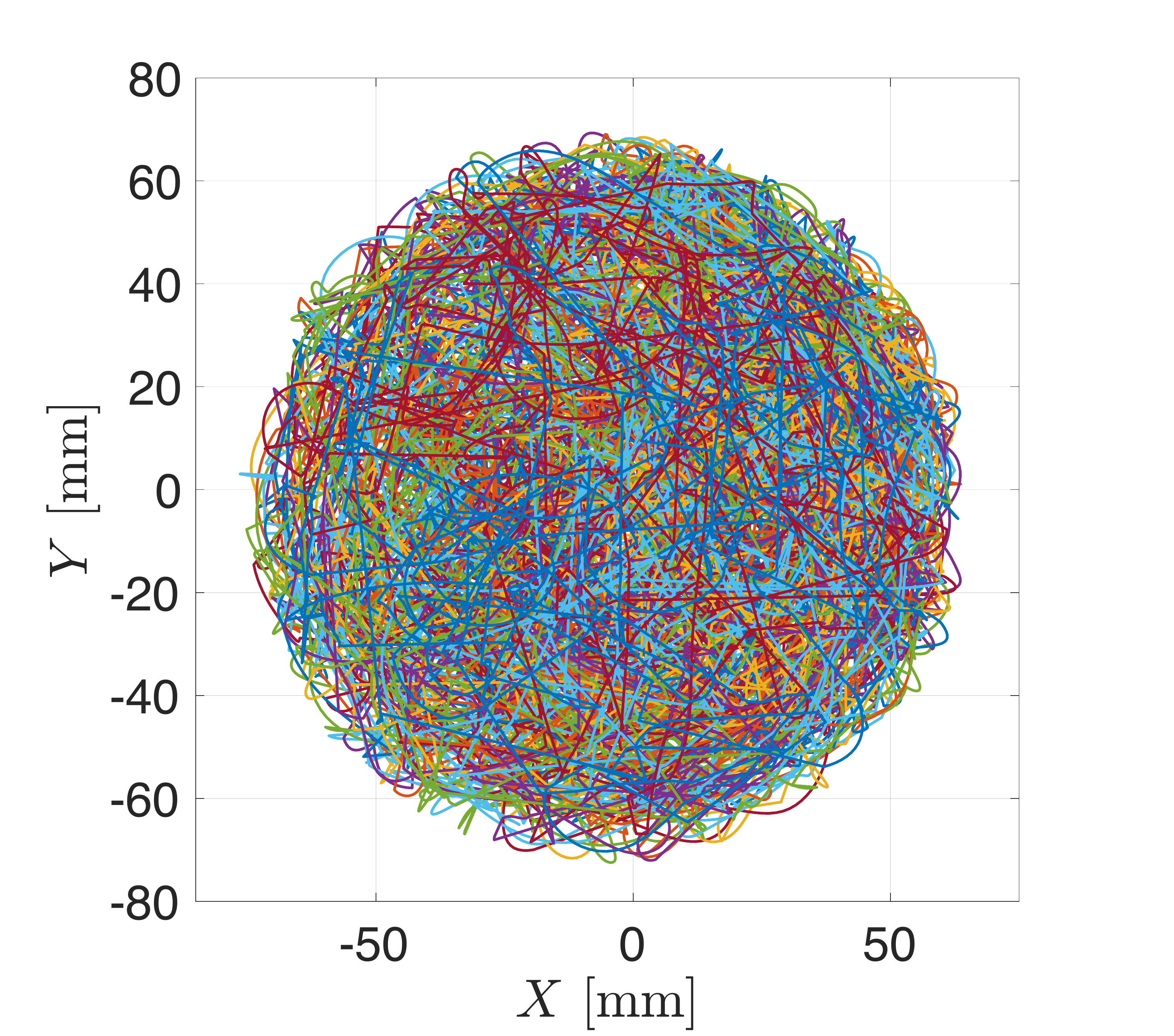}\\
				(a)\hspace{10cm}(b)
	\end{center}
\caption{(a) Top view of the experiment. 30 self-propelled camphor disks move at an air-water interface. Arrows represent the instantaneous velocity direction, colored according to the velocity magnitude. A movie is available as online supplementary material. (b) Superposition of the trajectories of the 30 particles over a five minute recording.  \label{fig:camphorDisks}}
\end{figure*}

The analogy between active matter and turbulence first emerged from qualitative observations of meso-scale patterns (such as whirls, jets and vortices) in dense biological suspensions~\cite{bib:mendelson1999_JBacteriol}, which seem ubiquitous in a wide variety of active systems~\cite{bib:bechinger2016_RevModPhys} and are visually reminiscent of typical structures of fluid turbulence. 
On a more quantitative level, multi-scale energy spectra were reported for {a few dense active systems}, for which a continuum approach allowed to define an {Eulerian flow field from the particles dynamics} \cite{Wensink_2012_jcm}, and several studies investigated how active systems can be described as hydrodynamical flows based on generalized Navier-Stokes equations with tunable non-linearity \cite{bib:toner1995_PRL,bib:wensink2012_PNAS,bib:bratanov2015_PNAS}. However such descriptions were found to {generally} exhibit flow spectra with non universal scaling exponents depending on model parameters. 
{For instance, a recent numerical and analytical study suggests that while displaying multi-scale energy spectrum, turbulence in active nematic systems falls into a universality class distinct from fluid turbulence, for which the energy is injected and dissipated at the same scale with no underlying cascade \cite{alert_arxiv}.}
{On the contrary, in a different class of active system made of small active spinners, experimental observations made on the generated fluid flow evidenced a kinetic energy spectrum compatible with Kolmogorov theory \cite{bib:kokot_pnas}, thus fostering the analogy. }

{To date, these various studies questioned such an analogy by focusing solely on} the spatial correlations (or spectra in Fourier space) computed from snapshots of an Eulerian velocity {field with such field {defined using the fluid if there is one, or as a continuous flow by local averaging over particles velocities in dense systems}. 
Yet turbulent flows both exhibit temporal and spatial fluctuations which are such that the time-scales measured in the Lagrangian framework are related by a specific scaling to the length-scales measured in the Eulerian framework \cite{bib:tennekes1972}. 
{Exploring the possible strong ties} between active systems and fluid turbulence {thus requires that we also look into the self-consistency of possible analogies} both in the Lagrangian and Eulerian frameworks. {Overall, this leaves the question whether classes of active systems can be described in the inertial turbulence framework essentially open.}

Finding such {active systems} exhibiting the same multi-scale dynamics as observed in turbulence would not only benefits to the future development of active systems but may also provide new insights for the longstanding description of turbulent flows. {This is the purpose of the present work to design and explore a synthetic active system displaying complex spatio-temporal dynamics amenable to combined Lagrangian and Eulerian characterization. The abiotic system considered is} made of a dilute assembly of $\mathcal{O}(30)$ self-propelled camphor swimmers which generate their own spatio-temporal fluctuations when confined in a box (Figure \ref{fig:camphorDisks}a). In the light of classical statistical indicators from turbulence, we analyze the dynamics of the swimmers (rather than the one of the fluid in between them as was done in \cite{bib:kokot_pnas}). Our experiments reveal striking similarities with fluid turbulence as both the Lagrangian and Eulerian dynamics of this active system are found, to a large extent, indistinguishable from that of fluid tracers in turbulence. In particular Eulerian statistics exhibit a well identifiable inertial range of scales following remarkably classical K41 scalings of fluid turbulence with a turbulent Reynolds number compatible with the time-scales measured in the Lagrangian framework.

%

\section{Experimental setup}
The scientific community has been prolific in the last decade in developing synthetic active particles and in investigating their individual and collective behaviors, with the goal to unveil generic physical properties of active matter~\cite{bib:bechinger2016_RevModPhys}. Here, we consider macroscopic synthetic swimmers based on the historical realization of camphor boats~\cite{bib:rayleigh1890_ProcRoySoc}. More precisely we use agar gel disks  loaded with precipitated camphor~\cite{bib:soh2008_JPhysChemB}, with radius $a=\SI{2.5}{\milli\meter}$ and height $h=\SI{0.6}{\milli\meter}$. When individually deposited at an air-water interface, the disks self-propel (with typical swim velocity $U_s$ in the order of \SI{10}{\milli\meter \per\second}) by {a} Marangoni effect arising from the camphor  spreading at the interface. {The individual hydrodynamical Reynolds number of such swimmers, $Re_p=U_sa/\nu$ (with $\nu={\SI{1e-6}{\meter\squared\per\second}}$ the kinematic viscosity of water), is of the order of 25, and no fluid turbulence is induced in {particles} wakes.}  
Figure~\ref{fig:camphorDisks}a shows a top view of our experiment, with 30 such self-propelled disks swimming on water in a \SI{140}{\milli\meter} diame\-ter Petri-dish. A video of the motion of the self-propelled disks is given as supplementary material. The motion of the self-propelled disks is recorded at a frame rate of 30 fps with a 1Mpx digital camera and their motion is analyzed using classical 2D particle tracking to reconstruct the Lagrangian trajectories of each individual camphor disk. Simultaneous trajectories about 5 minutes long are retrieved for all particles. Figure~\ref{fig:camphorDisks}b shows the superposition of the 30 trajectories simultaneously recorded, emphasizing that the available domain is explored ergodically. Velocity and acceleration of the particles are computed by convolution with first and second derivatives of a gaussian kernel \cite{bib:mordant2004_PhysicaD} to filter small scale noise.

\begin{figure*}[t]
	\begin{center}
		 \includegraphics[width=.45\textwidth]{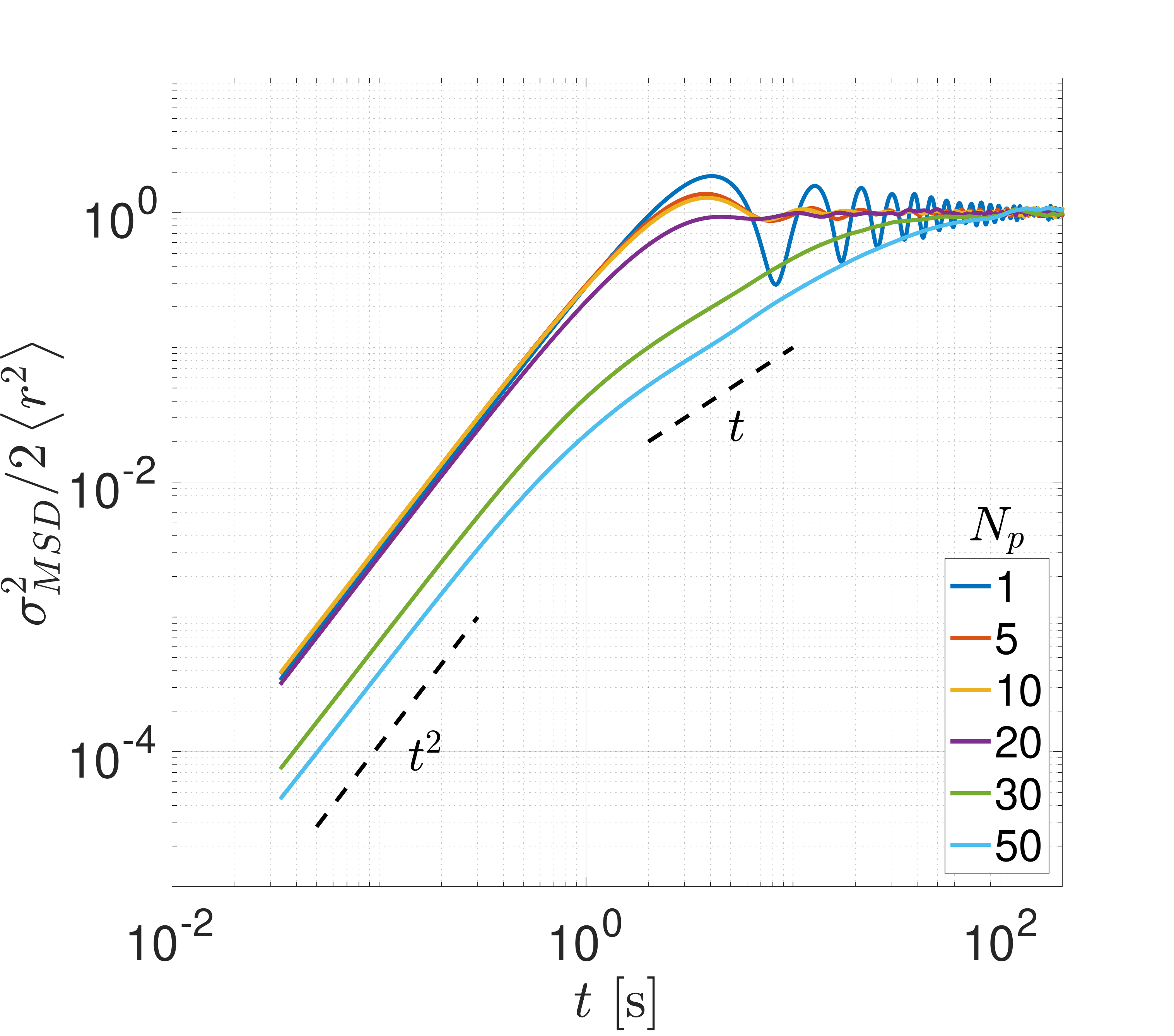}
		 \includegraphics[width=.45\textwidth]{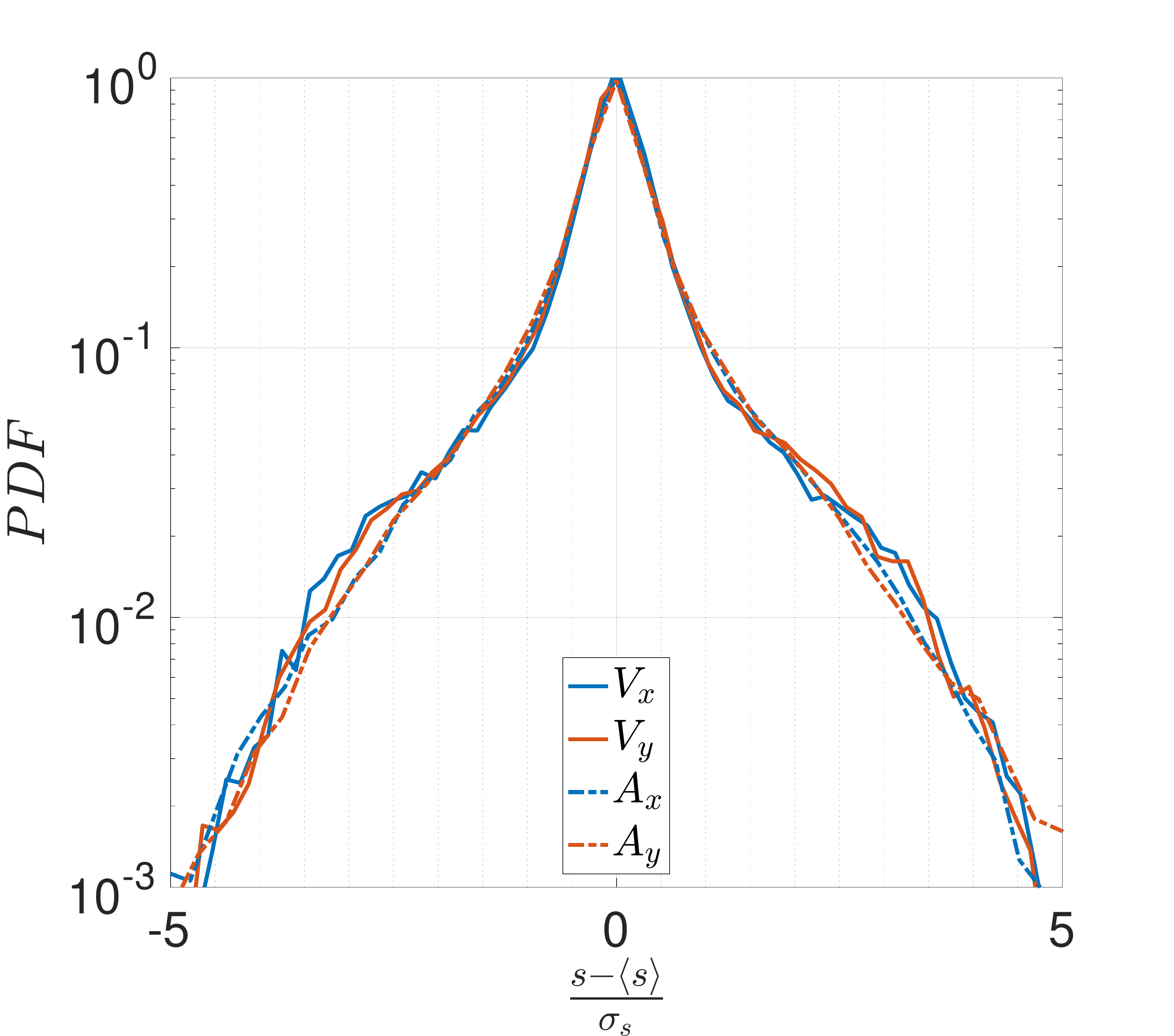}\\
		         (a) \hspace{10cm} (b)
	\end{center}
\caption{(a) Normalized mean square displacements ($\sigma_{MSD}^2/2\left<r^2\right>$, where $\left<r^2\right>$ is the average square radial position) of the camphor boats for experiments with different number of particles $N_p$. At large times, the MSD oscillates around and eventually saturates to one, due to the finite size of the experiment. For small numbers of particles (typically $N_p<20$) the MSD before saturation exhibit a dominant trivial ballistic motion (where the MSD grows as the square of time). For larger numbers of particles, an intermediate diffusive like regime (where the MSD grows linearly with time) appears revealing a non-trivial random dynamics with a finite correlation time-scale. (b) Probability density functions of velocity and acceleration components. Ought to the boundedness of the system, mean velocity and acceleration components are zero. The standard deviation of velocity components is $\sigma_{v_x}=\sigma_{v_y}=\left<v_i^2\right>_{i=x,y}^{1/2}=\SI{12.1}{\milli\meter\per\second}$ (corresponding to a root mean square velocity amplitude $\sigma_{|{v}|}=\SI{17.1}{\milli\meter\per\second}$) ; the standard deviation of acceleration components is $\sigma_{a_x}=\sigma_{a_y}=\left<a_i^2\right>_{i=x,y}^{1/2}=\SI{33.2}{\milli\meter\per\second\squared}$ (corresponding to a root mean square acceleration amplitude $\sigma_{|{a}|}=\SI{46.9}{\milli\meter\per\second\squared}$. \label{fig:MSDAll}}
\end{figure*}

In agreement with previous observations~\cite{bib:suematsu2015}, we find a transition between a dilute and {an interacting} swimming regime. In the dilute regime particles weakly affect each other, with individual trajectories  essentially straight except for short collision-like events or when a particle hits the boundaries. The {interacting} regime reveals on the contrary a much richer random collective dynamics with persistent high activity periods exhibiting large scale correlations, emblematic of the concept of active matter. This transition can be seen for instance from the mean square displacement (MSD) 
{of the camphor boats as a function of time $t$: 
\begin{equation}
\sigma_{MSD}^2(t)=\langle (\vec X(t+t_s)-\vec X(t_s))^2\rangle,
\end{equation}
where $\vec X$ is the instantaneous position of a camphor boat and $\langle \bullet \rangle$ an average over particles and starting time $t_s$ (see Figure~\ref{fig:MSDAll}a).
} For a small number of boats (typically $N_p<20$) the MSD exhibit two trivial regimes :  (i) at short times the MSD grows ballistically ($\sigma_{MSD}^2\propto t^2$) ; (ii) at long times the MSD oscillates around an asymptotic constant value. {These} regimes are consistent with a simple dynamics where particles essentially undergo a straight ballistic motion, with periodic re-orientation at the boundaries due to the finite size of the experiment. For a large number of particles ($N_p\gtrsim20$) {a} third intermediate regime appears, where the MSD grows diffusively ($\sigma_{MSD}^2\propto t$). This is characteristic of a randomization of the individual dynamics, with a finite correlation time scale, as observed for instance for {passive} particles following a {Langevin} dynamics and also {for tracers} in fluid turbulence, which is known to exhibit at large scales a similar effective turbulent diffusive behavior~\cite{bib:taylor1922,bib:sawford1991_PoFA}. Beyond the occurence of this dilute-to-{interacting} transition, the present article aims at further exploring the possible analogies between {interacting} active systems and fluid turbulence. The global behavior of the aforementioned transition for the collective dynamics of camphor boats have indeed already been addressed in previous studies \cite{bib:suematsu2015}, which have shown for instance that the transition occurs above a given threshold (setup dependent) of the total perimeter length $\ell_p=2 N_p \pi a$ and that the activity of the system then decreases (and eventually freezes for very dense systems). In our system, the dilute-to-{interacting} transition occurs around $N_p \approx 20$ (corresponding to a surface fraction of particles of $\phi_s \approx 2 \%$ and a total perimeter length $\ell_p\approx \SI{300}{\milli\meter}$), while the system is observed to freeze for $N_p\gtrsim 60$. The decrease of the individual activity when increasing the number of particles is also revealed in Figure~\ref{fig:MSDAll}a from the {shift of the different curves in the ballistic regime at increasing $N_p$ as the MSD is related to the velocity variance, $\sigma_v^2=\langle v_x^2\rangle + \langle v_y^2\rangle$, by the relation $\log \sigma_{MSD}^2 = \log \sigma_v^2 + 2 \log t$.} 

{The details of transition being outside the scope of the present work, we rather focus on unveiling the rich collective and multi-scale dynamics in the {interacting} regime ($N_p > 20$), which we will characterize at the light of statistical diagnoses borrowed from the turbulence community. As turbulence is a spatio-temporal phenomenon, correlation time-scales extracted in the Lagrangian framework (following fluid particle trajectories) are directly related to length-scales estimated in the Eulerian framework (from the analysis of the velocity field). We shall therefore study the statistics of the active particles both in the Lagrangian and Eulerian frameworks in order to explore possible quantitative analogies between the collective motion of the active disks and fluid turbulence.} The results discussed in the sequel were obtained with the number of particles fixed at $N_p=30$ 
which corresponds to a typical situation with rich collective dynamics and  strong individual activity.
%

%

\section{Lagrangian dynamics of active turbulence}
{We analyze in this section the Lagrangian dynamics of the present active system at the light of fluid turbulence.}
{As we shall see, characterization of relevant timescales will allow to determine an equivalent Reynolds number for our system, the consistency of which with Eulerian multi-scale spectra will be explored in the subsequent section.}\\
\paragraph{Single time statistics.}
Figure~\ref{fig:MSDAll}b shows the probability density functions (PDF) of the components of {the camphor boats} velocity $(v_x,v_y)$ and acceleration $(a_x,a_y)$ {where the variables have been centered --- although  the mean velocity and acceleration are vanishingly small ought to confinement of the system --- and reduced so that the shapes of the curves can be compared}. 
Several points can be noted : 
(i) the PDFs are identical for the two components, revealing the isotropy of the dynamics, 
(ii) PDFs of velocity and acceleration are identical.  {In the context of fluid turbulence such result would be interpreted as the absence of {\it internal} intermittency in the Lagrangian framework} {\footnote{{In the context of 3D (homogeneous and isotropic) turbulence the velocity PDF is close to Gaussian while the acceleration PDF is strongly non gaussian so that the PDF of velocity increments, $\delta_\tau v_i=v_i(t+\tau) - v_i(t)$ $(i=x,y)$, strongly depends on the time increment $\tau$. As acceleration is strongly linked to local dissipation, this is a consequence of {\it internal} intermittency (by contrast with {\it external} intermittency which refers to a bursting temporal dynamics often observed at a turbulent-non turbulent interface).}}}, (iii) PDFs are not Gaussian {(but without any stretched tail)}, a feature commonly reported in 2D fluid turbulence \cite{bib:bracco2000_PoF,bib:pasquero2001_JFM,bib:tsang2010_PoF}. 
Ought to the observed isotropy, statistics will be further investigated in the sequel considering only the $x$ component of the motion.\\

\begin{figure*}[t]
	\begin{center}
		\includegraphics[height=7cm]{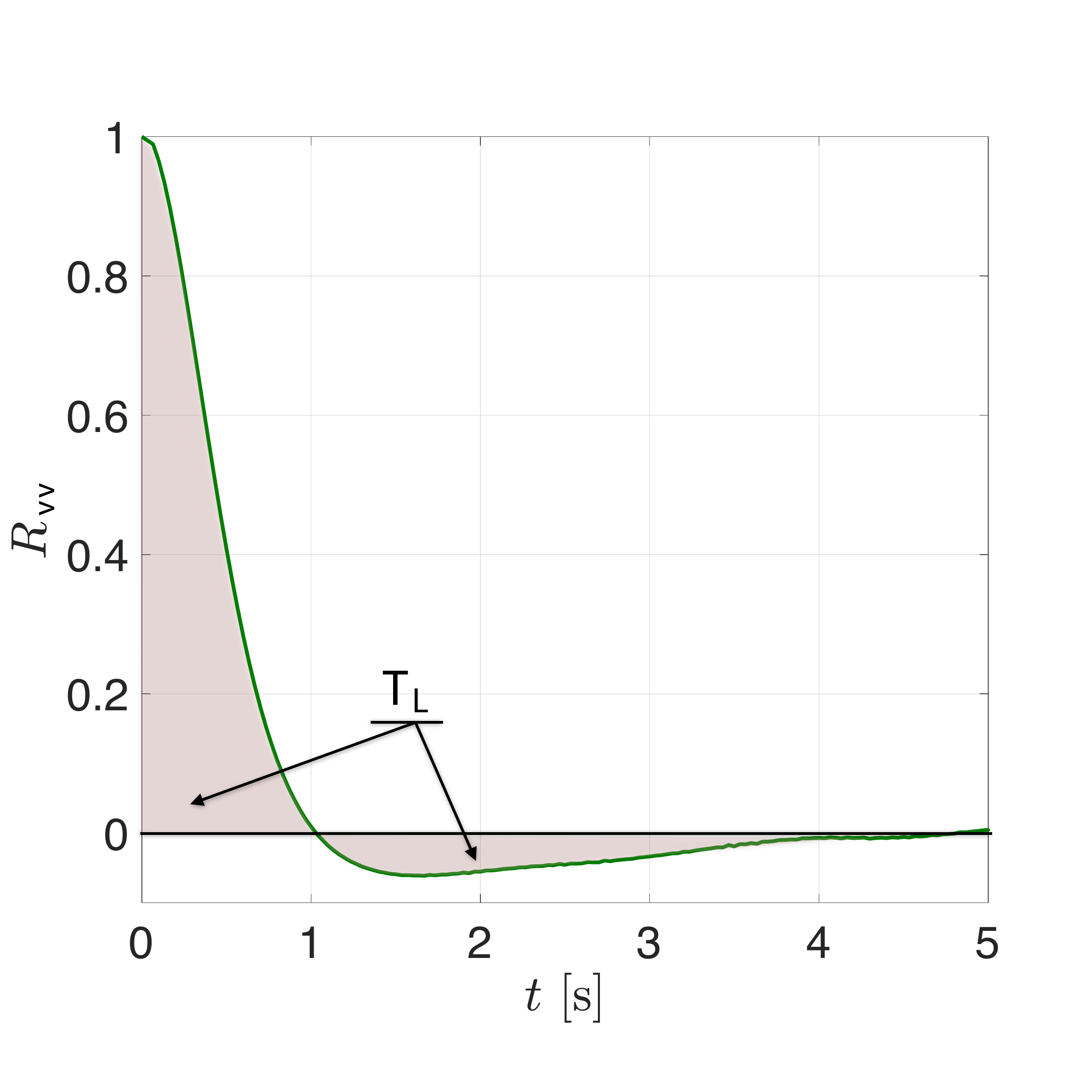}
		\hspace{2cm}
		\includegraphics[height=7cm]{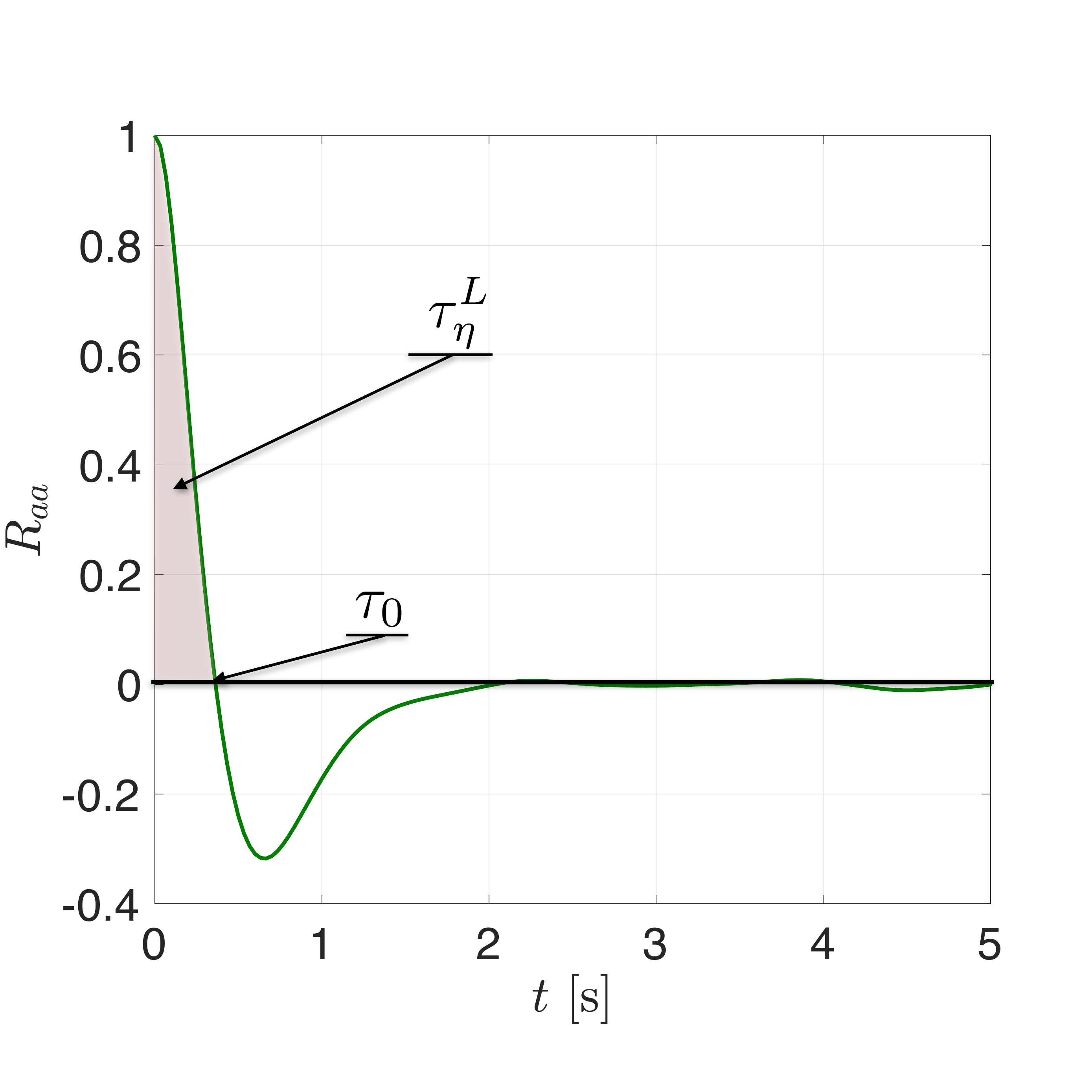}\\
				(a)\hspace{8.5cm}(b)
	\end{center}
\caption{Single particle - two times Lagrangian statistics. (a) Lagrangian autocorrelation function $R_{uu}$ of velocity component $v_x$. The light red colored area represents the estimate of the Lagrangian correlation time scale $T_L$. (b) Lagrangian auto-correlation function $R_{aa}$ of acceleration component $a_x$. $t_0$ corresponds to the shortest time such as $R_{aa}\left(t_0\right)=0$. The light red colored area represents the estimate of the Lagrangian dissipative time-scale $\tau_\eta^L$.}\label{fig:Ruu}
\end{figure*}

\paragraph{Two time{s} statistics.}

Besides the mean square displacement previously discussed, two of the most fundamental indicators used to explore the multi-scale Lagrangian dynamics of turbulence are the auto-correlation functions of velocity and acceleration components: 
{\begin{eqnarray}
R_{v_iv_i}(t)=\langle v_i(t_0)v_i(t+t_s)\rangle/\langle v_i^2\rangle\\
R_{a_ia_i}(t)=\langle a_i(t_0)a_i(t+t_s)\rangle/\langle a_i^2\rangle,
\end{eqnarray}
where again averages are taken over particles and starting time $t_s$.} These are important statistical tools allowing to quantify the multi-scale temporal dynamics of turbulence in terms of a hierarchy of relevant dynamical regimes: the Lagrangian dissipative regime at small scales where the dynamics is smooth and regularized by viscosity, the Lagrangian inertial regime at intermediate scales where the dynamics is rough and correlated, and the Lagrangian uncorrelated regime at large scales. The existence of an extended range of inertial scales is one of the most important features of fluid turbulence. In classical turbulence two important time-scales are defined, based on the correlation properties of velocity and acceleration, which delimit the extent of the inertial range: {(i) the integral Lagrangian time scale, which we define here as $T_L=\int_0^\infty \left| R_{vv} \right| \textrm{d}t$ (absolute values are requested here because the particle position is bounded \footnote{{Because the velocity signal $\vec v(t) = d\vec X/dt$ is statistically stationary with bounded position $\vec X(t)$, one has $\int_0^\infty R_{vv}(t') \textrm{d}t'=0$. This results is obtained from the equality $\textrm{d} \langle (\vec X(t) - \vec X(0))^2 \rangle/\textrm{d}t = 2 \int_0^t R_{vv}(t') \textrm{d}t'$.}})} and (ii) the dissipative Lagrangian time scale $\tau_\eta^L$, traditionally taken as~\cite{bib:calzavarini2009_JFM} $\tau_\eta^{L}=\int_0^{t_0} R_{aa}(t) \textrm{d}t$ (with $t_0$ the shortest time for which $R_{aa}(t_0)=0$). The dynamics occurring at scales smaller than $\tau_\eta^L$ is then referred to as \emph{dissipative}, the dynamics at scales larger than $T_L$ is referred to as \emph{uncorrelated} and the dynamics at scales intermediate between $\tau_\eta^L$ and $T_L$ is referred to as {\emph{inertial}}. $T_L$ and $\tau_\eta^{L}$ then define the Lagrangian Reynolds number~\cite{bib:sawford1991_PoFA} $Re^L=\left(T_L/\tau_\eta^L\right)^2$ which characterizes the extent of the Lagrangian inertial range of time scales.

We analyse here the multi-scale random Lagrangian dynamics of the active camphor discs within the same framework of fluid turbulence just described. Figures~\ref{fig:Ruu}a\&b show the velocity and acceleration auto-correlation functions for the active camphor disks. The corresponding integral and dissipative Lagrangian time scales are $T_{L}\approx\SI{0.6}{\second}$ and $\tau_\eta^{L}\approx \SI{0.2}{\second}$, leading to an equivalent Lagrangian Reynolds number $Re^L\approx 9$. Although this number may seem small at first sight, the Lagrangian Reynolds number is known in fluid turbulence to be significantly smaller than the classical integral Eulerian Reynolds number $Re^E=\left(L_E/\eta\right)^{4/3}$, where $L_E$ is the Eulerian (spatial) correlation length scale of the velocity fluctuations. Numerical simulations of fluid turbulence show that $Re^E$ and $Re^L$ are empirically related~\cite{bib:sawford1991_PoFA} by $Re^E\approx 88.6 Re^{L^{0.61}}$. In the fluid turbulence context a Lagrangian Reynolds number $Re^L\approx 9$ is therefore equivalent to an integral Reynolds number $Re^E\approx 340$. {To fix ideas, such conditions would correspond to a moderate turbulence, such as generated in a large scale (metric size test section) wind-tunnel blowing at \SI{3}{\meter\per\second} downstream a passive grid with \SI{7}{\centi\meter} mesh size \footnote{{Such flow corresponds to an integral Reynolds number $Re^E\approx 340$ and a Taylor scale Reynolds number $Re_\lambda \approx  70$, estimated through the relation $R_\lambda \approx \left(15 Re^E\right)^{1/2}$.}} \cite{bib:monchaux2010_PoF}.

{Overall, our active system of small (millimetric) camphor boats shows Lagrangian temporal signatures which are consistent with expectations for a typical fluid turbulence system. Based on this comparable multi-scale random dynamics, it is possible to determine the integral Reynolds number $Re^E\approx 340$ of the equivalent fluid system, which suggests a moderate turbulence regime. For a meaningful analogy to hold between the camphor interfacial swimmers and classical turbulence, it is mandatory that the spatial multi-scale dynamics of our active system be consistent with expected signatures in the very same turbulence regime. To that aim, we now turn to the study of Eulerian properties.}

%
%
%
%
%
%
%


\section{Eulerian dynamics of active turbulence}

In this section we investigate the Eulerian {multi-scale} dynamics of the camphor discs by considering the spatial correlations between their instantaneous velocity across the system rather than their temporal correlations along individual discs trajectories, as in the previous section. 
{According to the results of the Lagrangian dynamics, to match the properties of an equivalent fluid turbulence system ($Re^E\approx 340$; $R_\lambda\approx 70$), the camphor swimmers dynamics should} exhibit a narrow, but still visible inertial range of spatial scales. 
In particular the second order Eulerian structure function $S_2^E$ would exhibit a Kolmogorovian scaling ($S_2^E(r)\propto r^{2/3}$) over slightly less than one decade of scales $r$ (or equivalently the power spectral density would exhibit a narrow range of scales in Fourier space with a $k^{-5/3}$ regime). {Given the small number of active particles in the system, it is impossible to define an Eulerian velocity field by binning particles velocities in space as was done in Wensik {\it et al.} \cite{Wensink_2012_jcm} for dense suspensions. However such a field is not required for the computation of Eulerian structure functions of the velocity differences:
\begin{equation}
S_p^\parallel (r) = \left<\left|\frac{\left(\vec{v}_i(t)-\vec{v}_j(t)\right)\cdot\vec{r_{ij}}}{r_{ij}}\right|^p\right>, 
\end{equation}
where $r$ represents a bin of spatial scales, and the average is taken over all pairs $(i,j)$ of particles with separation $r_{ij}$ within that bin. Figure~\ref{fig:sn}a shows the longitudinal second order Eulerian structure function $S_2^\parallel (r)$ as a function of the separation between the active particles.} At small scales, $S_2^\parallel(r)$ vanishes for distances approaching the disk diameter (\SI{5}{mm}), as particles cannot interpenetrate. 
For large distances, $S_2^\parallel(r)$ tend to a constant asymptotic value $S_2^{\parallel\infty}\approx \SI{290}{\milli\meter\squared\per\second\squared}$ (which corresponds to $\sigma_{|v|}^2$, the variance of the velocity module) as expected for uncorrelated particles. 
At intermediate separations, a range of scales is observed, where $S_2^\parallel(r)\propto r^{2/3}$, reminiscently of a Kolmogorovian inertial range. This \emph{Kolmogorovian inertial range of active turbulence} is also visible in the Eulerian energy spectrum (estimated as the Fourier transform of the Eulerian correlation function:
\begin{equation}
R^{E}_{vv}(r)=1-\frac{S_2^\parallel(r)}{\sigma_{|v|}^2}, 
\end{equation}
shown in Figure~3b, which exhibits a noticeable $k^{-5/3}$ regime down to wave numbers corresponding to the particle diameter. We have also estimated higher order Eulerian structure functions $S_n^\parallel(r)$ (Figure~3c) which are found, within the range of \emph{inertial scales} just identified, to follow non-intermittent K41 inertial scalings, $S_n^\parallel(r)\propto r^{n/3}$. {Such result is consistent with the absence of {\it internal} intermittency that was observed in the Lagrangian framework when investigating velocity and acceleration PDFs.}


\section{Discussion}
\begin{figure*}
	\begin{center}

		\includegraphics[height=6cm]{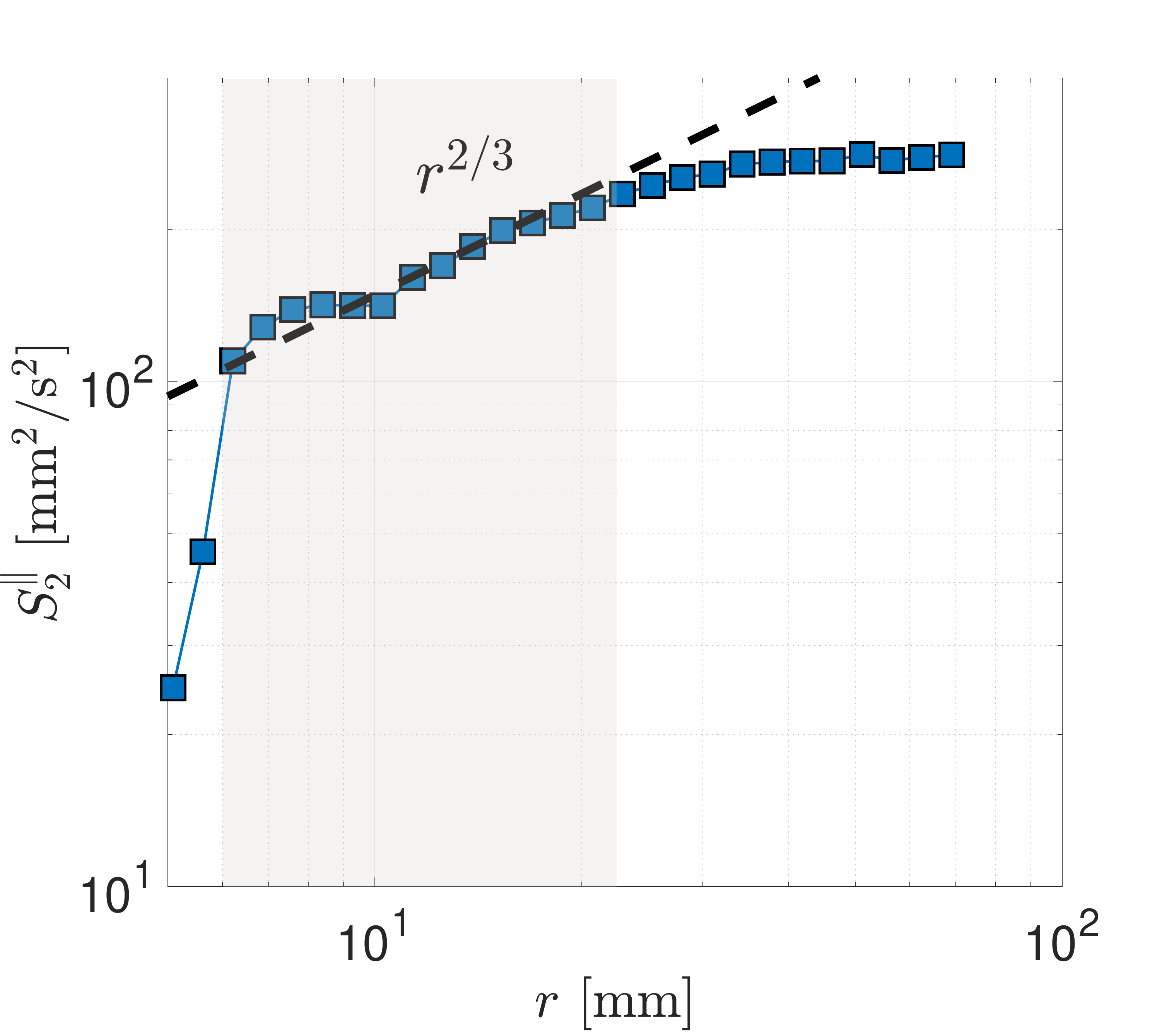}
		\hspace{2cm}
		\includegraphics[height=6cm]{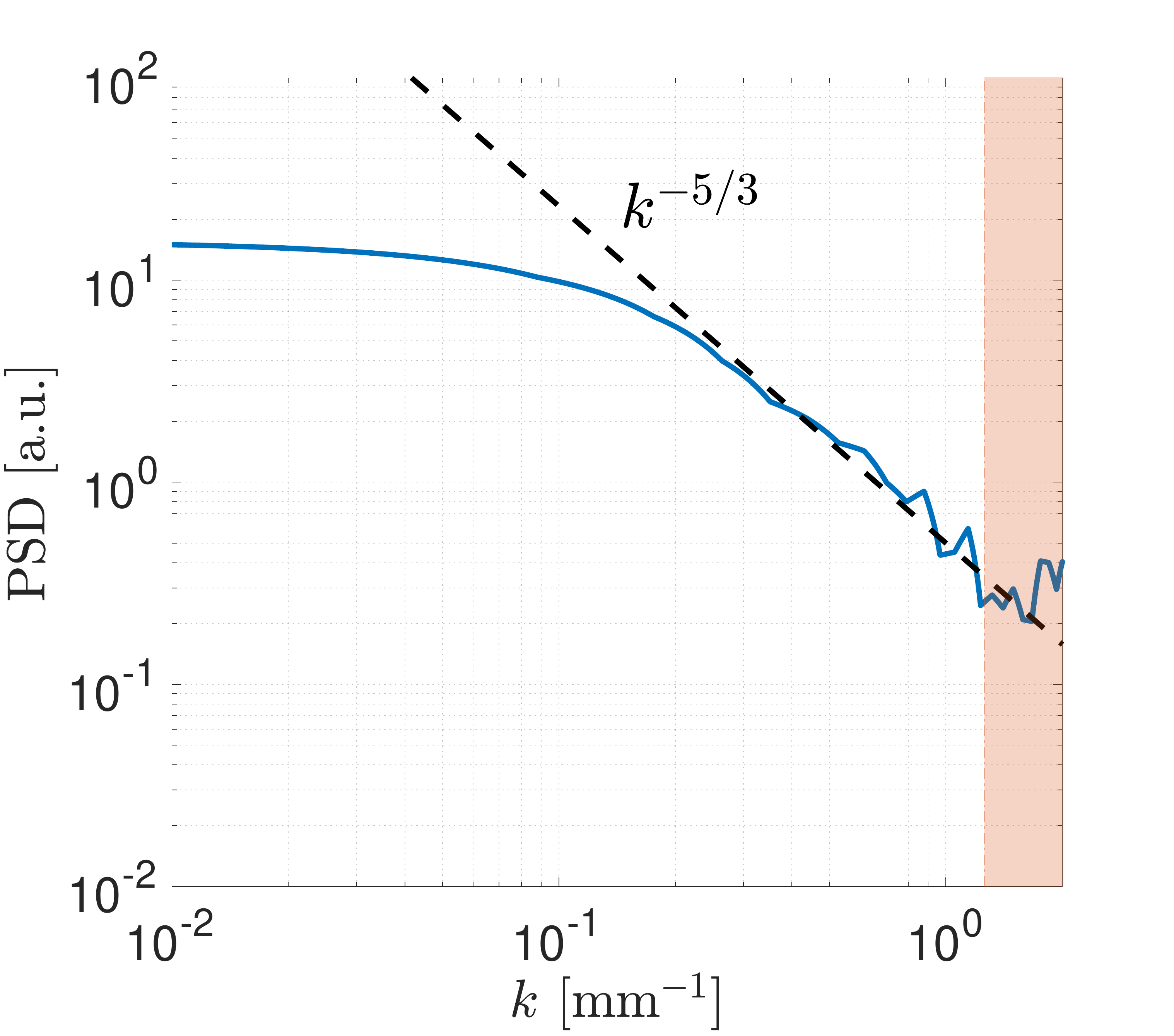}\\
		        	(a) \hspace{8cm} (b)\\

		\includegraphics[height=6cm]{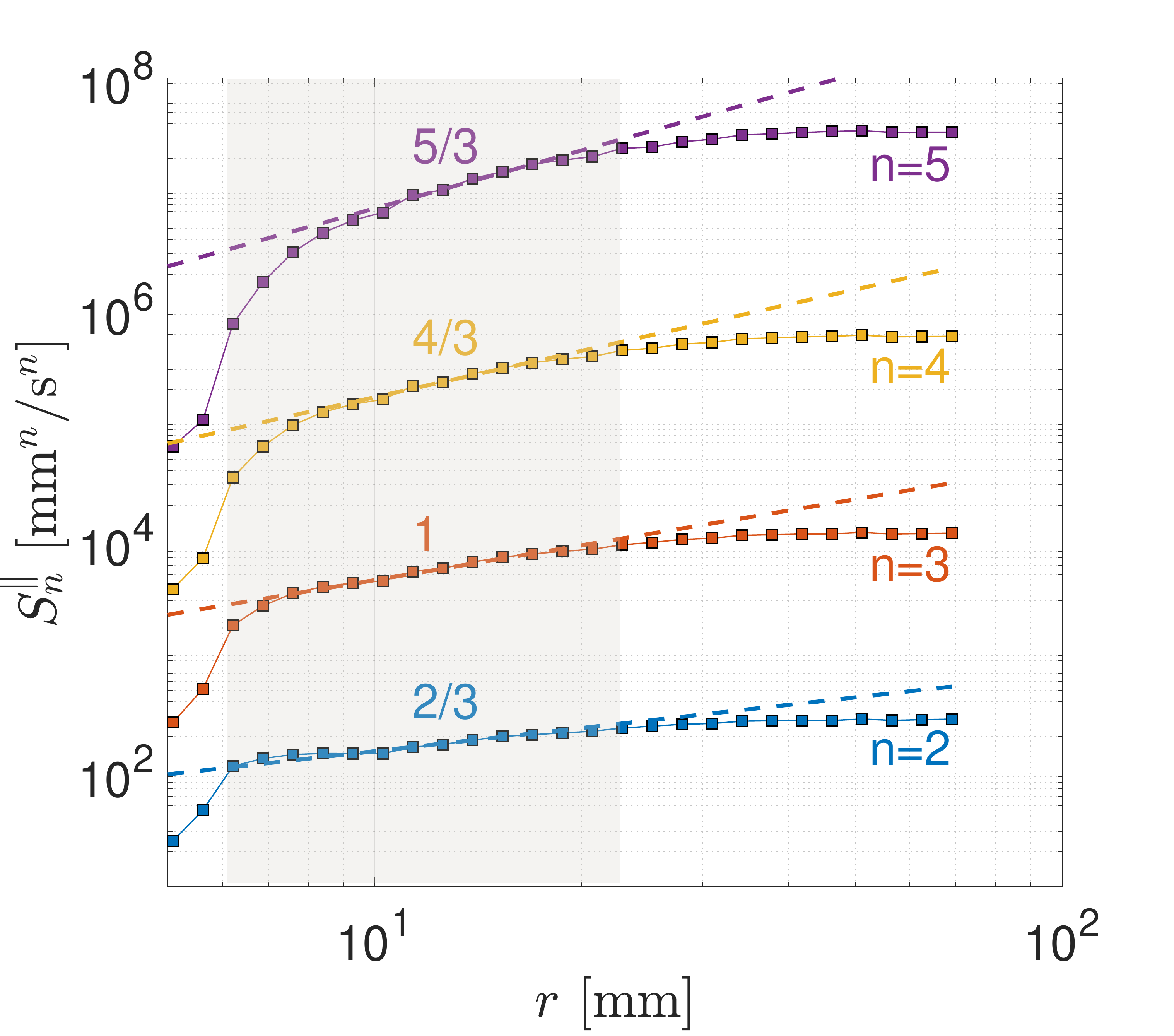}
		\hspace{2cm}
		\includegraphics[height=6cm]{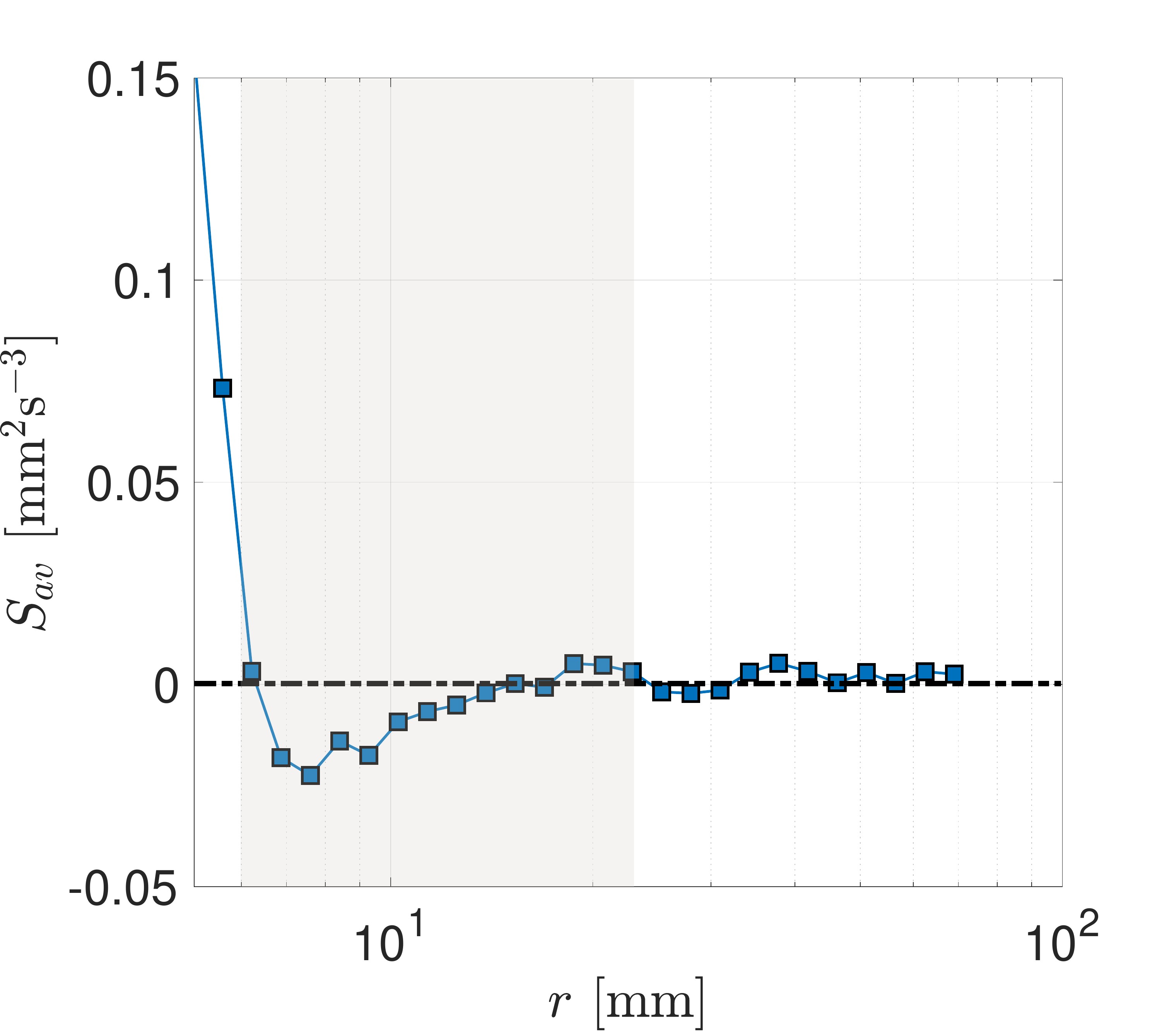}\\
		            	(c) \hspace{8cm} (d)\\
	\end{center}
\caption{Two particles - single time Eulerian statistics. (a) Second order longitudinal structure function $S_2^\parallel$ as a function of interparticle separation $r$. The dashed line represents the K41 prediction for fluid turbulence $S_2^\parallel\propto r^{2/3}$. (b) Power spectral density, obtained as the Fourier transform of the Eulerian correlation function. The dashed line represents the K41 prediction for fluid turbulence. The light red area on the right represents wave-numbers corresponding to scales smaller to the particles diameter. The Kolmogorov spectrum extends over almost one decade of scales, down to scales of the order of the particles diameter. (c) High order longitudinal structure functions $S_n^\parallel$ (for $n\leq 5$). Dashed lines indicate the corresponding non-intermittent K41 predictions ($S_n^\parallel \propto r^{n/3}$). The light gray area qualitatively indicate the corresponding \emph{inertial range} for which K41 scalings hold. (c) Crossed acceleration-velocity Eulerian structure function, whose amplitude in fluid turbulence is the twice the energy flux and the sign indicates the direction (direct or inverse) of the energy cascade : the negative sign here points towards a direct (from large to small scale) scenario.}
\label{fig:sn}
\end{figure*}

Altogether our results show remarkable quantitative analogies between the collective dynamics of interfacial active particles and fluid turbulence, both from the Lagrangian and the Eulerian points of view. The non-intermittent scalings found for the high order Eulerian structure functions are reminiscent of the inverse cascade regime in 2D fluid turbulence~\cite{bib:dubos2001_PRE,bib:boffetta2000_PRE}. Considering that in active matter energy is primarily injected at the particle level, it is therefore tempting to link the absence of intermittency to the existence of an inverse energy cascade of active turbulence. Figure~3d represents the energy flux {across} scales, as it could be estimated for fluid turbulence, based on the crossed velocity-acceleration Eulerian structure function~\cite{bib:jucha2014_PRL} $S_{av}^E(r)=\left<\delta_r\vec{a}\cdot\delta_r\vec{v}\right>$. The negative value of $S_{av}^E$ over the range of scales previously identified as \emph{inertial} reveals that the fluid-like energy cascade in the present system is actually not inverse, but direct (energy flows from large to small scales, as in 3D fluid turbulence, for which $S_{av}^E\sim -2\epsilon$). 
The multi-scale Eulerian dynamics {has} therefore not to be associated here {with} an upward energy flux originating from individual particles as energy source. It rather originates from large scale interactions, forcing down to smaller scales the collective  dynamics of the particles. In the present situation the chemical background of the dissolved camphor left behind the active particles could be the vector of such long range interactions. To support this scenario, it would be enlightening in future studies to monitor the chemical camphor background at the same time as the particles dynamics.

In the context of active matter, our results are important as they show the first example of active turbulence with inertial range dynamics quantitatively similar to a direct energy cascade in K41 fluid turbulence. To which extent this connection between active and fluid turbulence can be extended to other active systems, in particular with long range interactions (chemical, hormonal, visual, etc.), is an important opening of this work. 

In the context of fluid turbulence our results are also important as they provide a simple {experimental model} of a non-intermittent direct cascade of energy. The absence of intermittency may be related to the absence of a dissipative cut-off in the energy spectrum and structure functions, which follow K41 scalings down to scales of the order of the particle diameter. The small scale dynamics is therefore limited here by particle-particle collisions, but does not exhibit any effective viscosity behavior. 
In 2D turbulence, the absence of intermittency in the inverse cascade tends indeed to be associated to the idea that as energy flows upwards in scales, effect of viscous dissipation at small scales may be disregarded~\cite{bib:boffetta2000_PRE}. With this respect the parallel between active and fluid turbulence may help in the future to better understand the origin and the modeling of intermittency, which remains one of the biggest mysteries of turbulence in fluids.

\section{Acknowledgments}
This collaborative work was supported by the French research programs ANR-16-CE30-0028{, LABEX iMUST (ANR-10-LABX-0064), and IDEXLYON of the University of Lyon in the framework of the "Programme Investissements d'Avenir" (ANR-16-IDEX-0005).}

\bibliographystyle{plain}
\bibliography{main}

\end{document}